# Analyte-localization device for point-of-use processing of sub-millimetre areas on surfaces

Ali Oskooei[a] and Govind Kaigala[a]*

[a] IBM Research – Zurich, Säumerstrasse 4, 8803 Rüschlikon, Switzerland.

* E-mail: gov@zurich.ibm.com

## Highlights

1. A hand-held analyte-localization device operated with a mild level of applied vacuum.
2. The particular hydrodynamic design enables analyte localization on various substrates.
3. The device is fabricated by 3D printing, allowing for rapid, low-cost production.
4. The devised technology was shown to localize analytes effectively on substrates, cells blocks and human tissues sections.

## Abstract

We present a portable, simple-to-operate, point-of-use analyte surface localization device. We take advantage of a set of hydrodynamic design features and components that achieve passive analyte localization by means of a single vacuum input. The vacuum source can be supplied by mechanical or battery-operated vacuum sources that are portable and allow point-of-use operation in the absence of electricity. We discuss the governing hydrodynamic principle and design parameters in detail. In a case study, we demonstrate the applicability of our technology to successfully localize a solution of rhodamine on a polydimethylsiloxane (PDMS) substrate and produce sub-millimetre-sized spots via application of a mild vacuum pressure of less than 10 kPa. In addition, we demonstrate local staining of breast cancer cell blocks and on human breast cancer tissue sections.

## Abbreviations

DI: deionized
hALD: hand-held analyte-localization device
PDMS:  polydimethylsiloxane

# 1. Introduction

Surface biochemistry is critical in several areas pertaining to biomedical diagnostics[1-3], cell biology[4-6], and drug screening[7-9], for example. In conventional biochemistry laboratories, the entire biological sample is exposed to a reagent of interest to produce a certain result or to make a diagnosis (see figure 1A). In recent years, the concept of analyte localization has been introduced to perform, for example, multiplexed surface biochemistry with high spatial resolution[10]. With localization, only confined regions on the sample will be exposed to each reagent, resulting in reduced sample and reagent consumption, multiplexing potential, and enhanced mass transfer. In addition, high-resolution localization presents new possibilities such as single-cell studies (see figure 1B). Micropipettes[11, 12], nanopipettes[13-15], microfluidic probes[16-19], and scanning probe microscopy (SPM)-based instruments, such as atomic force microscope (AFM) dip-pen[2, 20], nanofountain probe[21, 22] and FluidFM[23-25], are all examples of technologies developed to perform localized micro- and nano-scale processing on substrates.

The localization techniques introduced to-date are technically sophisticated, and there is tremendous merit in their use within modern laboratory settings. However, they simply cannot be used as flexible, hand-held, point-of-use tools. Therefore, there remains a gap for the development of a portable, simple-to-operate, point-of-use analyte-localization technology. A major roadblock to achieving such a system is the need for sophisticated bulky motorized stages and flow-control systems, such as syringe pumps and pressure controllers, and electricity. Another obstacle is the complexity of operation and the need for expert skills to calibrate and operate the technologies. Furthermore, widespread use of such localization instruments is hampered by the cost and tedious microfabrication processes involved in their development. An ideal point-of-use localization technology must overcome the above-mentioned obstacles in order to become a simple, flexible and portable tool. The schematics in figure 1D describe and envision such a system. Of course, such a system can potentially also have a reduced resolution and precision compared with full-scale laboratory tools.

We present a hand-held analyte-localization device (hALD) that requires only a single vacuum input supplied through a portable vacuum source to operate. In designing the hALD, we have taken advantage of a set of hydraulic and hydrodynamic principles and features that enable analyte localization. The particular hydrodynamic design of the hALD and its processing apex enable semi-passive localization of an analyte on various surfaces upon contact with its apex, thus allowing flexible hand-held operation of the device without a need for precision motorized stages or other types of sophisticated holders. Furthermore, our devised technology is suited for manual operation on substrates of varying topology without the need for extremely flat or smooth surfaces or extensive alignment and calibration, a capability that has not been present in the previously-invented analyte localization techniques. Vacuum required for hALD operation can be supplied using mechanical or battery-operated sources that are portable, allowing point-of-use operation for field applications with no need for compressed air or electricity.

In this paper, we first introduce the working principle and design rules for our localization device, and then present analytical and numerical models describing the consumption rate and mass transport in hALD (section 2). We perform detailed hydrodynamic analysis of hALD operation, and describe variations in analyte-consumption rate as a function of the applied vacuum, the channel network geometry and the pressure head of the reservoir during operation. Using computational models, we demonstrate effective mass transport from hALD's apex to the substrate at all vacuum levels applied. In section 3, we validate our hALD technology experimentally: we describe the experimental procedures and showcase a representative application of hALD for surface localization and spotting. Surface spotting is done through localization of a solution containing rhodamine on a polydimethylsiloxane and latex substrates.

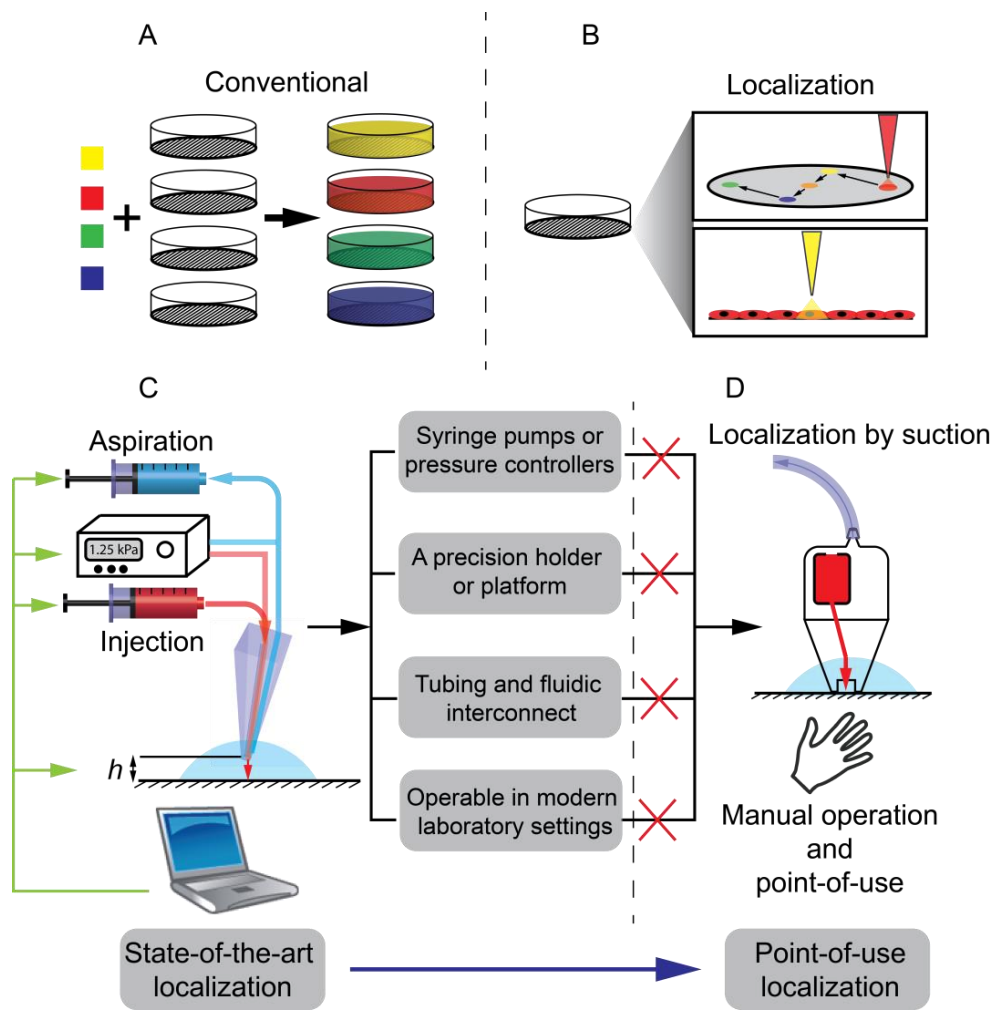

**Figure 1. Localized processing of biological substrate: present approaches versus point-of-use localization.** A) Conventional wet-bench biochemistry requiring the entire biological substrate for a specific test. B) Localization of multiple tests, such as multiple immunostaining using various biomarkers, performed on the same sample, thereby being conservative of samples and reagents. C) Current state-of-the-art localization techniques have been developed for localization, but require extensive infrastructure and accessories, largely rendering them usable only in modern laboratory settings and by trained laboratory personnel. D) View of an ideal localization technique that is simple to operate, portable and truly point-of-use, similar to pipettes used in life-sciences laboratories.

## 2. Design of the hALD

A rendered embodiment of our hand-held analyte-localization device is shown in figure 2A and B. As demonstrated in figure 2, the technology is self-contained with an integrated analyte reservoir and a vacuum port for supplying the negative pressure to localize the reagent at a recessed and textured processing apex (figure 2B). Within the hALD, there is a network of microchannels or capillaries (see figure 2C) that helps guide the analyte to the surface (injection line) and confine and remove the injected analyte and the immersion liquid from the surface (vacuum or aspiration line). The channel dimensions and elevation with respect to the surface have been optimized to ensure effective localization and analyte transport to the processing substrate.

The network of channels within the hALD is divided to four functional units: 1) The siphon channel is directly connected to the reservoir: the elevation difference must be designed to prevent unwanted analyte leakage while enabling flow from the reservoir when needed. 2) The recessed apex containing the injection and aspiration apertures (shown in figure 2B, bottom): six aspiration apertures distributed uniformly in the periphery of the injection aperture ensure uniform removal of the injected analyte and immersion liquid. 3) Eight-spoke-configuration apex channels designed to collect a surface immersion liquid. The radial inward flow of the immersion liquid helps sample localization. However, the localization will not be disrupted in the absence of an immersion liquid. In addition, the apex channels act as a release valve, attenuating the applied vacuum and helping to maintain a relatively steady consumption rate even in the presence of sharp vacuum fluctuations that occur when using manual or portable sources of vacuum. This passive flow-control ability sets the hALD apart from closed-channel systems, such as the continuous-flow microspotter (CFM)[26].  4) The vacuum or aspiration channel that removes the injected analyte and collected immersion liquid.

The operation protocol of the hALD comprises sample loading in the reservoir, flow initiation in the injection channel, and finally manual handling of the localized analyte on the substrate. Here, we analyze the hydrodynamics of the analyte in each stage of hALD operation, complete with equations and a mathematical description of the link between the pressure drop, the hydraulic head, and the flowrates.

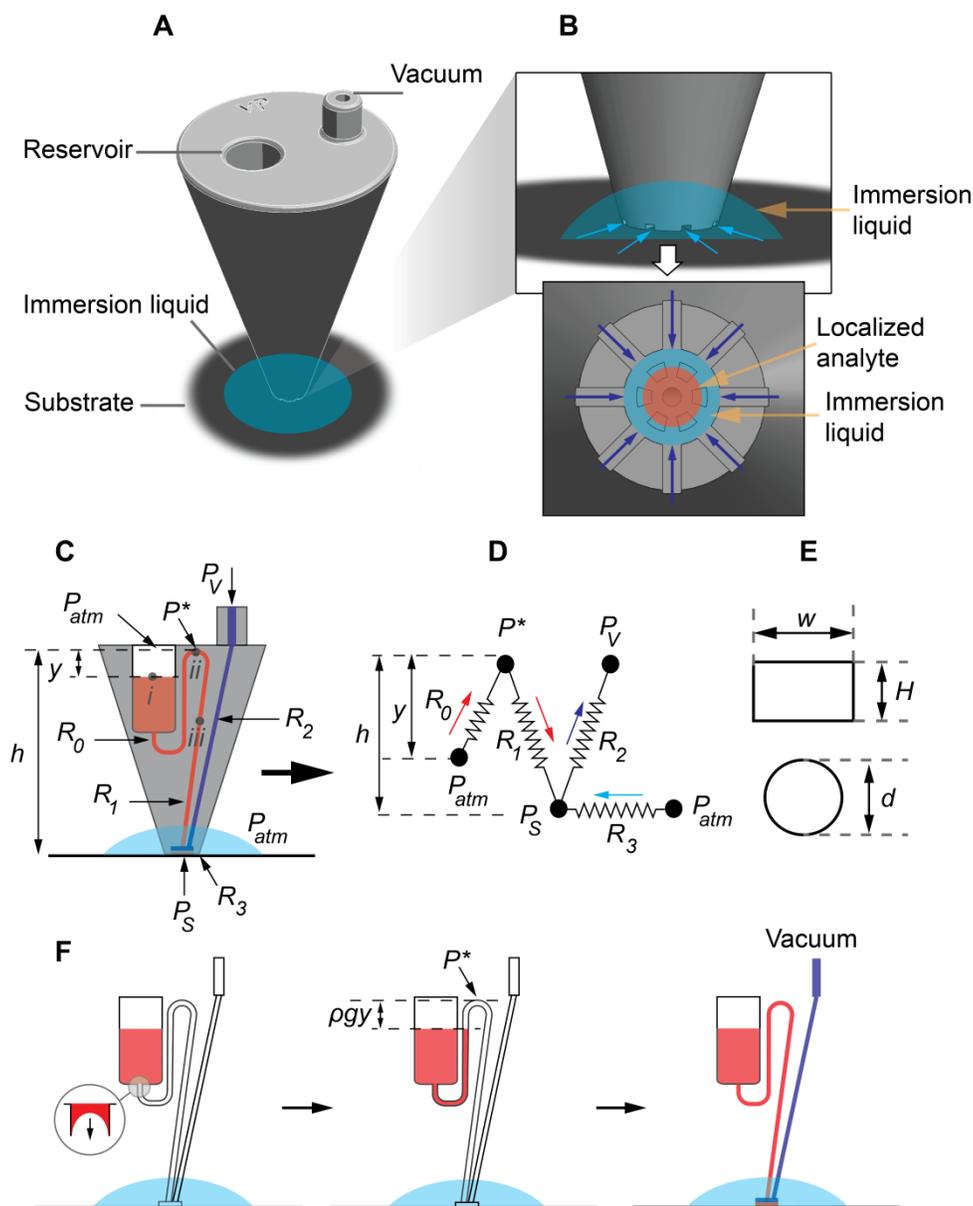

**Figure 2. Design and hydrodynamics of the hand-held analyte-localization device.** A) Illustration of the hALD, which consists of a reagent reservoir and a vacuum port connected to a vacuum source. B) Schematics of the hALD apex in contact with a substrate, localizing an analyte (shown in red) on the immersed substrate. Immersion liquid and the analyte from the reservoir meet at the apex of the hALD, localizing the analyte. The eight apex channels, shown in the schematic, collect the substrate-immersion liquid and form an analyte confinement on the substrate. C) The hydraulic design of the device consisting of a reservoir to store the analyte, an S-shaped siphon channel to prevent unwanted leakage from the reservoir, a vacuum channel that aspirates the injected analyte and immersion liquid, and a recessed apex containing the injection and aspiration apertures engulfed by apex channels that collect the immersion liquid (see figure 2B, bottom). D) Hydraulic circuit of the hALD showing the resistances and pressures that affect the hydrodynamic performance of the technology. E) Cross sections of the channel shapes used in the hALD design and their associated dimensions; the injection and vacuum channels are circular, whereas the apex channels are rectangular with a height of $H$ and a width of $W$. F) Evolution of the analyte flow from the moment the analyte is loaded into the reservoir until it is localized on a substrate. When the analyte is loaded into the reservoir, the analyte–air interface spontaneously enters the injection channel with wetting channel

walls, and proceeds further in the channel until it stops in the siphon S-channel. To move beyond this point, an applied external vacuum is needed, but once the interface flows beyond the S-channel, it will continue flowing even in the absence of an external vacuum source.

In the initial step of operation, the reservoir is filled with the liquid analyte, which is drawn into the reservoir capillary by interfacial forces between the liquid and the air (figure 2F, left). Without an externally applied pressure, the meniscus will advance in the capillary siphon (S-channel) until it reaches the hydraulic level in the reservoir (figure 2F, middle). At this point, the flow is stopped, preventing an inadvertent passive leakage of the analyte onto the surface. To drive the meniscus past this point and into the main injection channel of the hALD, a negative pressure of a magnitude equal to or greater than the difference in the pressure head between the reservoir and the highest point of the S-channel (i.e., $y$) is needed:

$$P^* \leq -\rho g y \,. \tag{1}$$

This implies that initially the vacuum applied should be sufficiently strong to induce a negative pressure smaller than $-\rho g y$ on the gas–liquid meniscus to overcome the pressure head in the S-channel and fill the injection channel. Once the entire injection capillary is filled with liquid, the liquid will continue to flow because of the hydraulic head, so that the requirement in equation (1) is no longer valid. At this stage, the hydraulic circuit in figure 2D applies, and the consumption flowrate from the reservoir towards the S-channel can be written as

$$Q_1 = \frac{\Delta P_0}{R_0} = \frac{-P^* - \rho g y}{R_0} \rightarrow P^* = -Q_1 R_0 - \rho g y. \tag{2}$$

We assume that the center of the hALD apex where the injection line, the vacuum line and the eight evenly distributed apex channels meet is small enough to be considered a point with a common pressure $P_s$. With this assumption, the flowrate for the injection line between the siphon and the apex can be calculated from the following relationship:

$$Q_1 = \frac{\Delta P_1}{R_1} = \frac{P^* - P_s + \rho g h}{R_1} \rightarrow P_s = P^* + \rho g h - Q_1 R_1. \tag{3}$$

For the vacuum line between the apex and the vacuum port, we can obtain the flowrate as

$$Q_2 = \frac{\Delta P_2}{R_2} = \frac{P_s - P_v - \rho g h}{R_2} \rightarrow P_s = Q_2 R_2 + P_v + \rho g h \,. \tag{4}$$

The flow of immersion liquid drawn from outside of the apex through the eight apex channels can be determined from the following equation:

$$P_s = -Q_3 R_3 \rightarrow Q_3 = \frac{-P_s}{R_3} \,. \tag{5}$$

Note that fluid flow in the above equations is assumed to be directed towards the vacuum line. From the law of conservation of mass, we can determine the following relationship between the flowrates in the hydrodynamic circuit of the hALD:

$$Q_2 = Q_1 + Q_3 \,. \tag{6}$$

For a known applied vacuum pressure (i.e., $P_v$) at the vacuum port of the hALD, we can organize the governing hydraulic equations as

$$
\begin{cases}
P_s + Q_3 R_3 = 0 \\
P_s - Q_2 R_2 = P_v + \rho g h \\
P^* + Q_1 R_0 = -\rho g y \\
P^* - P_s - Q_1 R_1 = -\rho g h \\
Q_1 - Q_2 + Q_3 = 0
\end{cases} . \tag{7}
$$

Equation (7) can be presented in matrix form as

$$
AX = b \rightarrow
\begin{bmatrix}
0 & 0 & R_3 & 1 & 0 \\
0 & -R_2 & 0 & 1 & 0 \\
R_0 & 0 & 0 & 0 & 1 \\
-R_1 & 0 & 0 & -1 & 1 \\
1 & -1 & 1 & 0 & 0
\end{bmatrix}
\begin{bmatrix}
Q_1 \\
Q_2 \\
Q_3 \\
P_s \\
P^*
\end{bmatrix}
=
\begin{bmatrix}
0 \\
P_v + \rho g h \\
-\rho g y \\
-\rho g h \\
0
\end{bmatrix}. \tag{8}
$$

Solving the above system of equations analytically, the hydrodynamic characteristics of the hALD are determined. From equation (8), the solution for the consumption rate, $Q_1$, as a function of the hydraulic resistances and the applied vacuum ($P_v$) is derived as:

$$
Q_1 = \frac{\rho g (h - y) - \frac{(P_v + \rho g h)}{(1 + R_2 / R_3)}}{R_0 + R_1 + \frac{1}{(1/R_2 + 1/R_3)}} . \tag{9}
$$

Equation (9) suggests a linear relationship between the rate of consumption and the applied vacuum and Therefore an increase in rate of consumption increase with the the applied vacuum ($P_v$).

To design and optimize the dimensions of the flow channels, all hydraulic resistances (i.e., $R_1$, $R_2$, and $R_3$) in (8) and (9) must be quantified in terms of the geometrical dimensions of hALD. The hydraulic resistance of microchannels has been widely studied, and analytical and empirical relationships are available in the literature. In our design procedure, we calculated the resistance for rectangular channels with width $W$, height $H$ and length $L$ from the following relationship described in detail elsewhere[27]:

$$
R = \frac{12 \mu L}{W H^3 \left(1 - 0.63 \frac{H}{W}\right)} , \ H < W, \tag{10}
$$

For circular channels with diameter $d$ and length $L$, we determine the resistance as

$$
R = \frac{128 \mu L}{\pi d^4} . \tag{11}
$$

Combining equations (8), (10) and (11) and iteratively solving the system of equations, we determined the flowrates and pressures within the hALD for a wide range of channel dimensions, and selected a desirable set of dimensions that matched our flowrate requirements. Plots of the consumption rate versus applied vacuum ($P_v$) for various dimensions of the injection channel ($d_1$) and the vacuum channel ($d_2$) and the apex height ($H$) are presented in figure 3A-C. Based on the design calculation, we adopted a hALD design with $d_1 = 500$ µm, $d_2 = 300$ µm and $H = 200$ µm. In addition, for the hALD design adopted, we quantified the flowrate variations induced by variations in the analyte level in the reservoir as analyte is being consumed during operation under a constant vacuum (i.e., variable $y$, constant $P_v$). The flowrate variation due to analyte consumption is plotted in figure 3D for various magnitudes of applied vacuum.

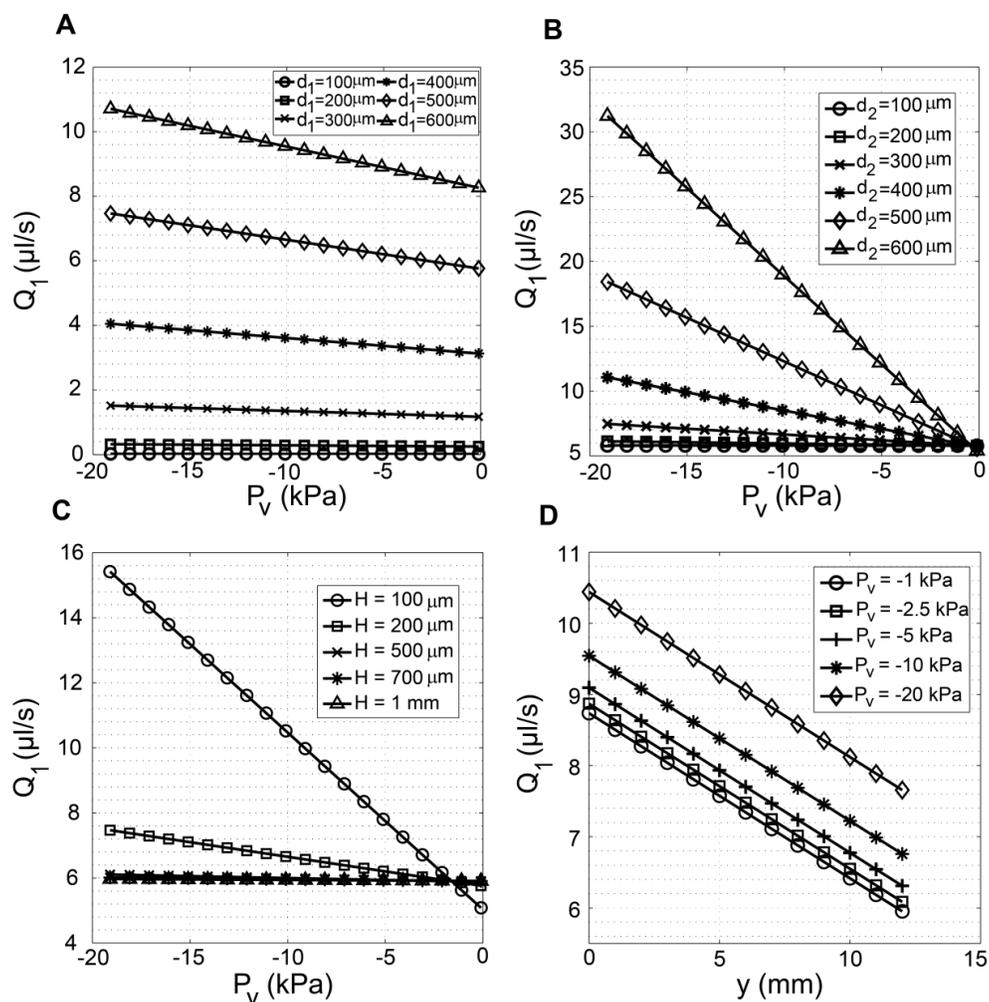

**Figure 3. Analytically-derived characterization of operational parameters.** (A) Analytically obtained analyte consumption rate plotted against the applied vacuum for various diameters of the injection channel while keeping the dimensions of the vacuum channel and the apex channels constant ($d_2 = 300$ μm and $H = 200$ μm). As shown in the plot, the consumption rate increases with a greater slope than the injection channel diameter does. (B) Consumption rate plotted against applied vacuum for various vacuum channel diameters, $d_2$, while maintaining the dimensions of the injection and apex channels constant ($d_1 = 500$ μm and $H = 200$ μm). With increased channel diameter (i.e., decreased resistance), the rate of consumption rises more sharply (i.e., exhibits a steeper slope). (C) Consumption flowrate plotted against applied vacuum pressure for various apex channel heights, $H$, while maintaining the dimensions of the injection and vacuum channels constant ($d_1 = 500$ μm, $d_2 = 300$ μm). As shown in the plot, the consumption rate at large channel heights is defined by the pressure head of the analyte in the reservoir and is independent of the applied vacuum. D) Variation in consumption rate as the analyte is being consumed and the analyte level in the reservoir drops (i.e., $y$ grows). As expected, the consumption rate drops for all applied vacuum levels as the analyte in the reservoir is being consumed.

A 3D axisymmetric numerical simulation of the hALD was performed in ANSYS Multiphysics to ensure that the analyte remains both localized and in contact with the substrate under various applied vacuum levels and will not flood the substrate or detach from the surface, which would result in a loss of effective surface processing. The results of the numerical models are presented in figure 3B. The analysis is performed for three different apex heights ($H = 200\ \mu m$, $500\ \mu m$ and $1000\ \mu m$). As shown in figure 4, the analyte will flood the surface for all apex heights when the vacuum level is not sufficient to localize the analyte effectively. Beyond a threshold vacuum level ranging from $P_S = -2.5\ Pa$ to $-5\ Pa$, the analyte is effectively localized as shown by the sharp step in the concertation distribution on the substrate. The numerical results demonstrate that for all three designs, analyte detachment will not occur even at very high vacuum levels ($P_v = -1\ atm \approx -100\ kPa$). This inherent lack of sensitivity to the applied vacuum levels will ensure a robust performance of hALD, which will not be disrupted by fluctuations in the applied vacuum as long as a required minimum level of vacuum is maintained ($P_S = -2.5\ Pa$ to $-5\ Pa$).

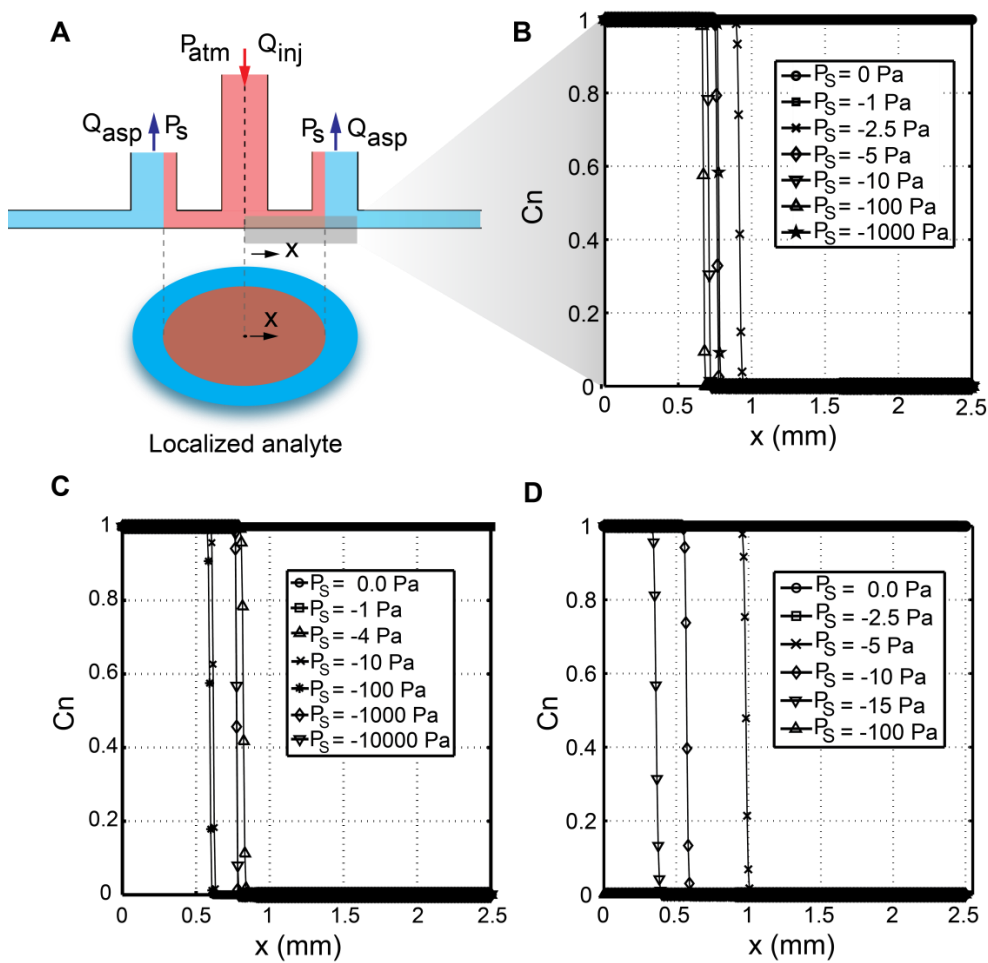

**Figure 4. Computational study of effective analyte contact with and species transfer to the substrate at various apex recession depths, $H$, for our hALD design with $d_1 = 500\ \mu m$ and $d_2 = 300\ \mu m$.** (A) Schematic illustrating the hALD apex, with the localized analyte shown in red. (B) Normalized IgG concentration (Cn) distribution ($D = 4.5 \times 10^{-11}\ m^2/s$) on a substrate produced by hALD with an apex recession of $H = 200\ \mu m$ for various apex pressures, $P_S$. Effective localization occurs for apex pressures below $-2.5\ Pa$, corresponding to an applied vacuum of $P_v = -2300\ Pa$ (from equation (7)). The radial distance from center of the apex is denoted by X. (C) Normalized concentration distribution for hALD with $H = 500\ \mu m$, showing a similar trend as (B) with effective analyte localization and surface contact for

pressures below $P_S = -4$ Pa, corresponding to $P_v = -31.80$ kPa. (D) Normalized concentration plots of hALD having an apex recession of $H = 1$ mm. The effective localization and contact are established at $P_S = -5$ Pa (~ $P_v = -39.3$ kPa), but contact with the surface is lost for apex pressures lower than $P_S = -15$ Pa as obtained using computational models. However, for this value of $H$, $P_S = -15$ Pa is not achievable, even at $P_v = -100$ kPa, which is the lowest achievable vacuum level (from equation (7)).

# 3. Experimental

In this section, we demonstrate experimentally the effectiveness and utility of our technique for analyte localization on surfaces and for substrate patterning. Moreover, we discuss device fabrication and operation as well as the materials and methods adopted in performing the experiments.

## Material and methods

The devices were designed in CAD software and 3D-printed (3D Labs, St. Georgen, Germany) in an acrylic photopolymer. Thin layers of polydimethylsiloxane (PDMS) and of latex were used as the substrates for analyte-localization and substrate-patterning experiments. PDMS and latex were appropriate choices of substrate material as they can readily be stained through contact with a solution containing rhodamine[28, 29]. To visualize the localized analyte and to stain and form spots on the PDMS and latex substrates, 0.5 mM solutions of rhodamine and fluorescein in deionized water were used as the analyte. Deionized water was used as the immersion liquid covering all substrates tested. The hALD was connected to a vacuum source via 0.25" rubber tubing connected to the vacuum port of the hALD without fittings or adapters. The applied vacuum was moderate in strength and maintained between zero and −10 kPa using a vacuum gauge to represent the vacuum level producible with small, portable vacuum sources. Human breast tumor tissue slides (ProSci Incorporated, Poway, CA, USA) and MCF7 breast cancer cell line blocks (CellMax™ FFPE Control Cell Line Block, AMS Biotechnology Ltd, Bioggio-Lugano, Switzerland) were used in cell and tissue staining experiments with hALD. A hematoxylin solution (Aqueous Hematoxylin, GeneTex, Inc., Irvine, CA, USA) was used to stain tissue sections and cell blocks.

## Results and discussion

Our hALD experiments focused on two main objectives: first, to verify complete and effective localization of the injected analyte through visualization of the analyte confinement at the hALD's apex and secondly, to study and verify transport and the effective surface contact of the injected analyte by producing rhodamine spots on PDMS and latex substrates.

In accordance with the design concept of the hALD, the experimental setup was simple and portable. The hALD was connected to vacuum via rubber tubing, and the rhodamine solution was fed into the reservoir using a pipette. Immersion liquid (i.e., DI water) was supplied to the substrate to keep the processing surface immersed at all times. An image of the experimental setup consisting of the hALD connected to the vacuum via tubing is shown in figure 5A. The schematic in figure 5B describes the two-step operation procedure of the hALD: the liquid analyte is loaded into the reservoir, and the vacuum is applied to localize the analyte on the substrate. The simple operation procedure renders the hALD readily applicable by laboratory personnel with minimal training or technical or engineering knowhow. The complete operation procedure of the hALD for staining a thin layer of latex with an aqueous solution of rhodamine is shown from beginning to end in supplementary video 1.

A fluorescent microscopy image of the localized fluorescent analyte in contact with the substrate is reproduced in figure 5C. In addition, a fluorescent microscopy video of the localized fluorescein solution in contact with a glass substrate is available as supplementary video 2. As demonstrated in supplementary video 2, the localization of analytes using the hALD was stable and robust despite manual operation and the movements and instabilities that inherently occur during manual operation of instruments.

To demonstrate successful surface processing and patterning using the hALD, we performed rhodamine staining experiments on a PDMS layer as substrate. The resulting stained footprint is shown in figure 5D. The stained spot is irregular in shape owing to the hand-held nature of the operation that inevitably induces minor vibrations and movements in the apex of the device, resulting in a movement of the analyte confinement and the irregular shape of the footprint. However, for applications requiring surface bio-patterning and diagnostics, such as immunoassays or immunohistochemistry, the shape of the footprint is not a critical factor and will not alter the end results. As shown in figure 5D, the footprint size in the hALD is predominantly defined by the dimensions and the arrangement of the injection and aspiration channels. In our hALD design, it was nearly 1 mm as intended by our choice of channel size and distribution. However, it is possible to achieve various footprint sizes as needed for different applications by performing design optimization using the governing hydrodynamic equations in (7).

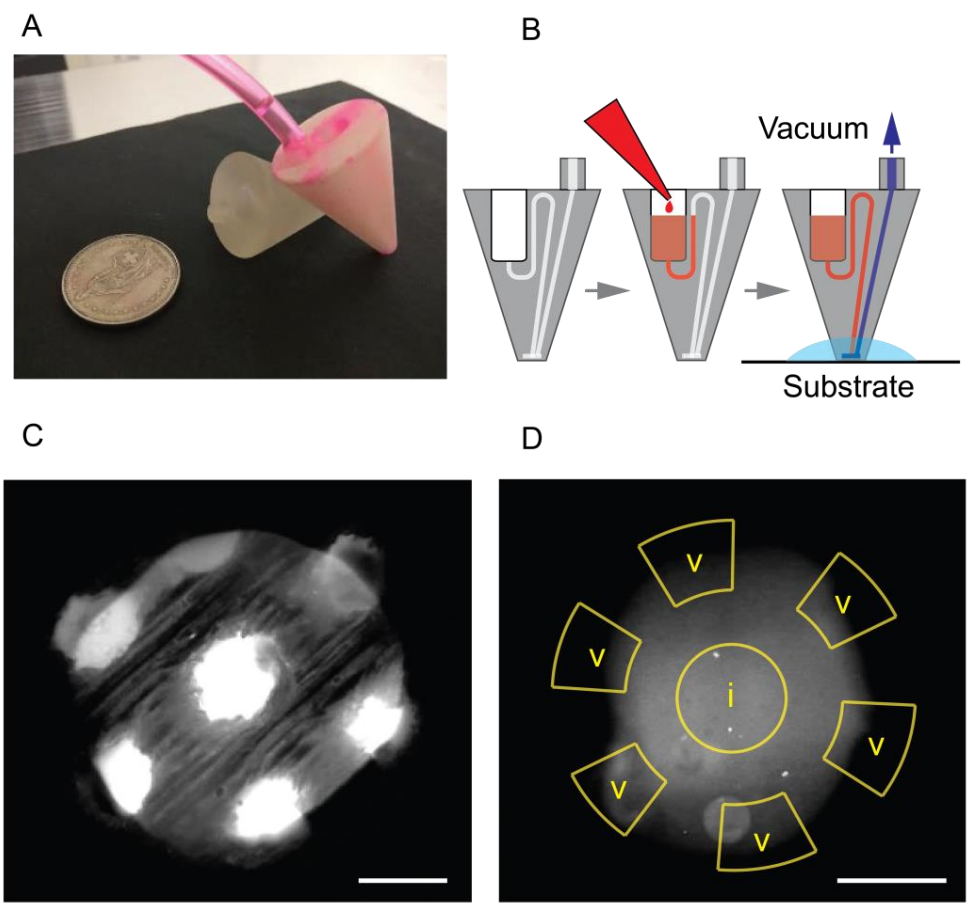

**Figure 5. Experimental demonstration of the hALD function.** A) Image of our hALD consisting of a reagent reservoir and a port connected to a vacuum via rubber tubing. B) Schematic illustrating the operational steps of hALD: reservoir loading followed by application of vacuum and, as a result, analyte localization on the substrate. C) Fluorescent image of localized fluorescein solution in contact with a glass substrate, with water as immersion liquid. The immersion liquid and the fluorescein solution from the reservoir were drawn into the vertex of the hALD, thereby localizing the fluorescein solution. D) A spot formed on the surface of a PDMS substrate by localization of a rhodamine solution on the surface for duration of 1 min using the hALD. In the apex sketch, "i" stands for the injection channel and "v" represents a vacuum channel. Scale bars represent 500 μm.

To demonstrate the utility and applicability of hALD with biological samples, we performed hematoxylin staining experiments using hALD and two sets of biological samples: human breast tumor tissue sections and MCF7 breast cancer cell-line blocks. The schematic in figure 6A illustrates the experimental procedure: to produce a single hematoxylin spot, hALD reservoir was loaded with 100 μl of hematoxylin solution and the vacuum level was set to ensure an analyte consumption rate of nearly 3 μl/s which resulted in reservoir depletion within 30 seconds. The hALD was then placed and held manually on the tissue or cell-block samples, immersed in water, for a duration of 30 seconds. Thereafter, the hALD was removed and the stained samples were imaged under a microscope. Figure 6B-E demonstrate the hematoxylin staining results. Figure 6B and C present the stained breast tissue sample under 4× and 20× magnification respectively while figure 6D and E demonstrate the stained cell-block results. As shown in figure 6, hALD successfully confined hematoxylin on both the tissue and cell samples and created a confined stained spot with no contamination to the surrounding areas or physical damage to the sample itself. The results in figure 6 confirm that hALD can be used safely and effectively for performing rapid point-of-use localized biochemistry on biological samples.

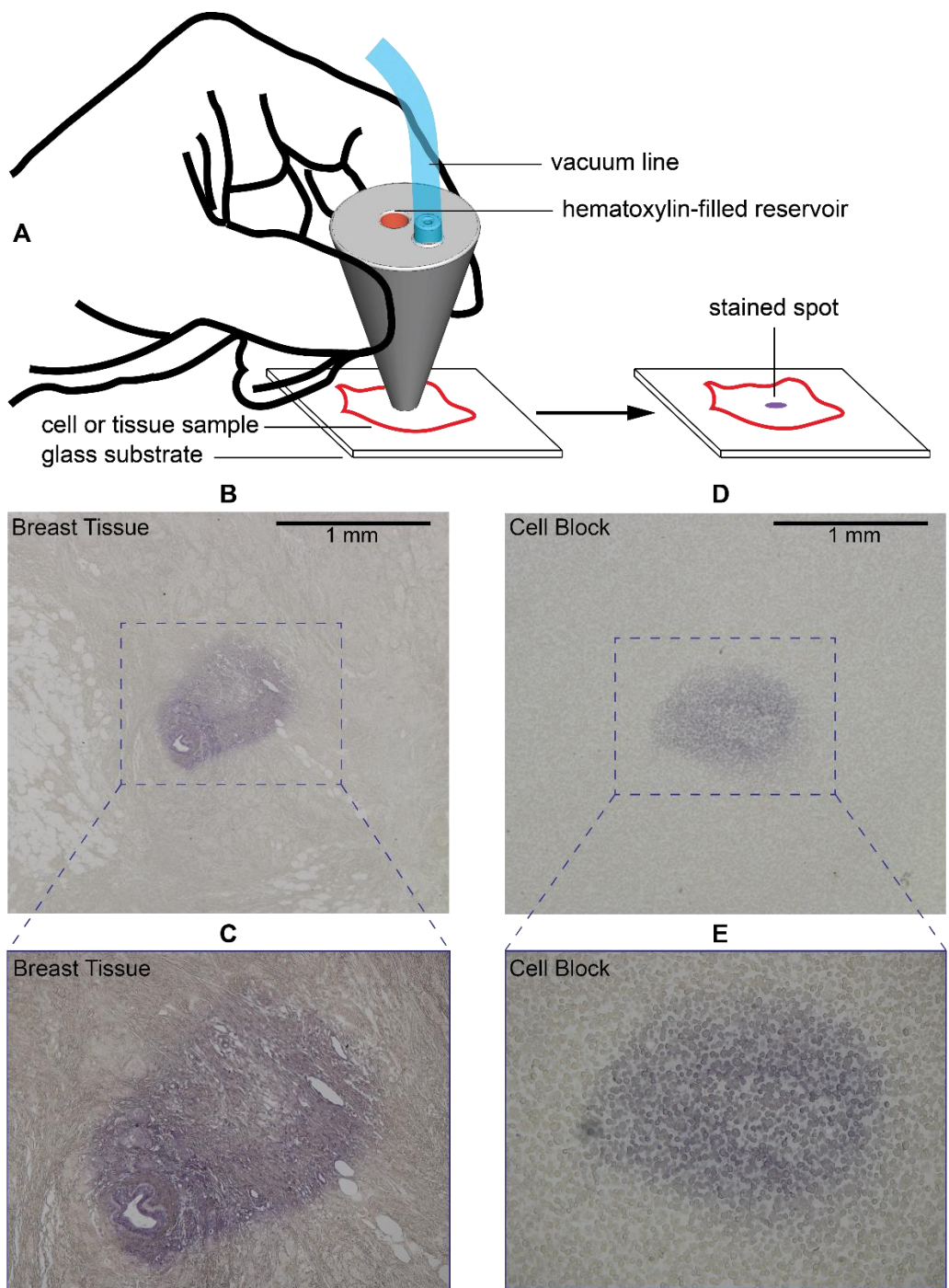

**Figure 6. Application of hALD for cell and tissue staining.** A) The schematic describing the experimental procedure for the staining of cell and tissue samples with hematoxylin using hALD B) A 4x magnified brightfield image of a human breast tumor tissue sample stained with hematoxylin using hALD C) A 20x

magnified brightfield image of the same tissue sample demonstrating the stained cells and ducts within the breast tissue section. D) A 4× magnified brightfield image of a MCF7 cell-block stained with hematoxylin using hALD. E) A 20× magnified brightfield image of the same MCF7 cell-block demonstrating the stained cells.

## 4. Concluding remarks and outlook

We introduced a hand-held analyte-localization device that is suitable for point-of-use applications thanks to its passive design that can be operated with only a low-level applied vacuum and maintains a relatively uniform consumption rate. We presented the hydrodynamic design of the technology and outlined key features that ensure robust performance in the surface localization of analytes. We confirmed and validated our analytical findings with numerical simulations that demonstrated complete and effective surface localization of analytes using our choice of design configuration. We demonstrated analyte localization using our hand-held device through fluorescent-flow visualization of localized fluorescent analytes. Furthermore, we demonstrated successful surface staining using our hALD technology by creating sub-millimetre-sized rhodamine spots on PDMS and latex surfaces. We showcased hALD utility for performing point-of-use localized biochemistry on biological samples by locally staining breast tissue sections and cell-blocks with a hematoxylin solution using hALD. We envision our device to offer new opportunities for performing localized biochemistry in both laboratory and field settings with low-level battery-powered or manual vacuum sources as the only requisite equipment for effective analyte localization. The hand-held design concept and ease of operation pave the way for the hALD to be used as a common laboratory tool for localized surface processing. The potential versatility of the hALD is further reinforced by its ease of fabrication in polymeric materials using readily available techniques such as 3D printing. We envision that achieving footprint sizes below 50 µm with hALD technologies will become possible as 3D-printing technology evolves.

### Acknowledgements

This work was supported by the European Research Council (ERC) Starting Grant, under the 7[th] Framework Program (Project No. 311122, BioProbe). We thank Anna Fomicheva for help with cell and tissue sample preparation and David Taylor for advice on 3D printing. Emmanuel Delamarche and Walter Riess are acknowledged for their continuous support.

### Biographies

**Ali Oskooei** is a researcher with IBM Research Zurich currently working on computational methods for cancer diagnostics. He completed his PhD at the University of Toronto where he developed technologies related to lab-on-a-chip micro-devices that have been patented or published in prestigious scientific journals. His research interests involve combining mathematical and computational modeling and experimentation to develop novel biomedical diagnostic devices and solutions.

**Govind V. Kaigala** is a Research Staff Member at IBM Research- Zurich and is currently leading activities on liquid-based non-contact scanning probe technologies – microfluidic probe – and is championing concepts on "microfluidics in the open space" and "tissue microprocessing". His broader research interests include leveraging micro- and nanosystems for microchip-based chemical & biomolecular analysis applied to medical applications. Prior to joining IBM, he was a postdoctoral fellow in the Mechanical Engineering and Urology at Stanford University. He received his Ph.D and M.Eng from the University of Alberta, Canada. He has authored and co-authored 44 scientific publications in leading journals, 60 conference papers, and about 25 patent families, and editor of a book. He is the recipient of several IBM awards, including a Research Division Accomplishment Award in 2014, the Horizon Alumni Award from the University of Alberta, and he is a Senior Member of IEEE.

# References

1.  K.-B. Lee, E.-Y. Kim, C. A. Mirkin and S. M. Wolinsky, *Nano Letters*, 2004, **4**, 1869-1872.
2.  K.-B. Lee, S.-J. Park, C. A. Mirkin, J. C. Smith and M. Mrksich, *Science*, 2002, **295**, 1702.
3.  R. Dunlap, *Immobilized Biochemicals and Affinity Chromatography*, Springer Science & Business Media, 2013.
4.  J. Li, W. Zhao, R. Akbani, W. Liu, Z. Ju, S. Ling, C. P. Vellano, P. Roebuck, Q. Yu and A. K. Eterovic, *Cancer Cell*, 2017, **31**, 225-239.
5.  B. G. Keselowsky, D. M. Collard and A. J. García, *Biomaterials*, 2004, **25**, 5947-5954.
6.  U. Sauer, *Sensors*, 2017, **17**, 256.
7.  E. M. Forsberg, C. Sicard and J. D. Brennan, *Annual Review of Analytical Chemistry*, 2014, **7**, 337-359.
8.  B. R. Stockwell, *Nature*, 2004, **432**, 846-854.
9.  R. Akbani, K.-F. Becker, N. Carragher, T. Goldstein, L. de Koning, U. Korf, L. Liotta, G. B. Mills, S. S. Nishizuka and M. Pawlak, *Molecular & Cellular Proteomics*, 2014, **13**, 1625-1643.
10. G. V. Kaigala, R. D. Lovchik and E. Delamarche, *Angewandte Chemie International Edition*, 2012, **51**, 11224-11240.
11. A. Ainla, G. D. M. Jeffries, R. Brune, O. Orwar and A. Jesorka, *Lab on a Chip*, 2012, **12**, 1255-1261.
12. A. Ainla, E. T. Jansson, N. Stepanyants, O. Orwar and A. Jesorka, *Analytical Chemistry*, 2010, **82**, 4529-4536.
13. T. Takami, B. H. Park and T. Kawai, *Nano Convergence*, 2014, **1**, 1-12.
14. A. Bruckbauer, P. James, D. Zhou, J. W. Yoon, D. Excell, Y. Korchev, R. Jones and D. Klenerman, *Biophysical Journal*, 2007, **93**, 3120-3131.
15. K. Jayant, J. J. Hirtz, I. Jen-La Plante, D. M. Tsai, W. D. De Boer, A. Semonche, D. S. Peterka, J. S. Owen, O. Sahin and K. L. Shepard, *Nature Nanotechnology*, 2016.
16. D. Huber, J. Autebert and G. V. Kaigala, *Biomedical Microdevices*, 2016, **18**, 40.
17. G. V. Kaigala, R. D. Lovchik, U. Drechsler and E. Delamarche, *Langmuir*, 2011, **27**, 5686-5693.
18. D. Juncker, H. Schmid and E. Delamarche, *Nature Materials*, 2005, **4**, 622-628.
19. A. Oskooei and G. Kaigala, *IEEE Transactions on Biomedical Engineering*, 2016. DOI: 10.1109/TBME.2016.2597297
20. R. D. Piner, J. Zhu, F. Xu, S. Hong and C. A. Mirkin, *Science*, 1999, **283**, 661-663.
21. W. Kang, F. Yavari, M. Minary-Jolandan, J. P. Giraldo-Vela, A. Safi, R. L. McNaughton, V. Parpoil and H. D. Espinosa, *Nano Letters*, 2013, **13**, 2448-2457.
22. K.-H. Kim, N. Moldovan and H. D. Espinosa, *Small*, 2005, **1**, 632-635.
23. O. Guillaume-Gentil, E. Potthoff, D. Ossola, C. M. Franz, T. Zambelli and J. A. Vorholt, *Trends in Biotechnology*, 2014, **32**, 7, 381-388.

24. P. Stiefel, T. Zambelli and J. A. Vorholt, *Applied and Environmental Microbiology*, 2013, **79**, 4895-4905.

25. A. Meister, M. Gabi, P. Behr, P. Studer, J. Vörös, P. Niedermann, J. Bitterli, J. Polesel-Maris, M. Liley, H. Heinzelmann and T. Zambelli, *Nano Letters*, 2009, **9**, 2501-2507.

26. K. A. Smith, B. K. Gale and J. C. Conboy, *Analytical Chemistry*, 2008, **80**, 7980-7987.

27. M. Tanyeri, M. Ranka, N. Sittipolkul and C. M. Schroeder, *Lab on a Chip*, 2011, **11**, 1786-1794.

28. S. A. Pfeiffer and S. Nagl, *Methods and Applications in Fluorescence*, 2015, **3**, 034003.

29. R. Samy R, T. Glawdel T, C. Ren, *ASME 2008 First International Conference on Micro/Nanoscale Heat Transfer, Parts A and B*:257-265. doi:10.1115/MNHT2008-52269.